\documentclass[a4paper,12pt]{article}
\usepackage[utf8]{inputenc}
\usepackage{amsmath}
\usepackage{amssymb}
\usepackage{bbold}
\usepackage{amsfonts,bm}
\usepackage{geometry}
\usepackage{fullpage}
\usepackage{graphicx}
\usepackage{caption}
\usepackage{todonotes}
\usepackage{multicol}
\usepackage{subcaption}
\usepackage{slashed}
\usepackage{float}
\usepackage{color,xcolor}
\usepackage{epstopdf}
\usepackage{units}
\usepackage[normalem]{ulem} 
\usepackage{jheppub} 
\usepackage{physics}
\usepackage{bm,paralist,xspace,url,relsize}
\usepackage[]{hyperref}

\title{Exploring the potential of FCC-hh to search for particles from $B$ mesons}
 \author[a]{Alexey~Boyarsky,}
 \author[a]{Oleksii~Mikulenko,}
 \author[a]{Maksym~Ovchynnikov,}
 \author[b]{and Lesya~Shchutska}
  \affiliation[a]{Instituut-Lorentz, Leiden University, Niels Bohrweg 2, 2333 CA Leiden, The Netherlands}
 \affiliation[b]{Institute of Physics,
High Energy Physics Laboratory,
École Polytechnique Fédérale de Lausanne (EPFL),
CH-1015, Lausanne, Switzerland}

\emailAdd{boyarsky@lorentz.leidenuniv.nl}
\emailAdd{mikulenko@lorentz.leidenuniv.nl}
\emailAdd{ovchynnikov@lorentz.leidenuniv.nl}
\emailAdd{lesya.shchutska@epfl.ch}
\date{}

\begin{document}
\abstract{The Future Circular hadron Collider (FCC-hh) is a proposed successor of the Large Hadron Collider (LHC). FCC-hh would push both the energy and intensity frontiers of searches for new physics particles. In particular, due to higher energy and luminosity than at the LHC, at FCC-hh there would be produced around $\simeq\!30$ times larger amount of $B$ mesons and $\simeq 120$ times of $W$ bosons, which then may decay into feebly interacting particles. In this paper we demonstrate the potential of FCC-hh by studying its sensitivity to heavy neutral leptons (HNLs) with masses $m_{N}<m_{B}$. We consider various locations of a displaced decay volume embedded in the planned infrastructure of FCC-hh. We demonstrate that FCC-hh may substantially improve the reach of the parameter space of HNLs as compared to the searches proposed at the LHC.}

\maketitle

\section{Introduction}
\label{sec:introduction}
The Standard Model (SM) of particle physics is extremely successful in describing accelerator data. However, it fails to explain several phenomena: neutrino oscillations, dark matter, and baryon asymmetry of the Universe. One way to resolve the incompleteness of the SM is to extend it by adding renormalizable interaction operators with SM particles and new feebly interacting particles. These hypothetical particles may explain the phenomena by themselves, or instead serve as a ``portal'' to the dark sector~\cite{Alekhin:2015byh}. 

Among the portals, an attractive option is heavy neutral leptons (HNLs) -- hypothetical fermion particles coupled to the Higgs doublet $H$ and the left-chiral SM lepton doublet $L_{\alpha}$, $\alpha = e,\mu,\tau$ (see~\cite{Abdullahi:2022jlv} and references therein):
\begin{equation}
    \mathcal{L}_{\text{HNL}} = F_{\alpha I}\bar{L}_{\alpha}\widetilde{H}N_{I}+\text{h.c.},
    \label{eq:hnls}
\end{equation}
where $N_{I}, I = 1,2,\dots$ is an HNL, $\widetilde{H}$ is the Higgs doublet in the conjugated representation, and $F_{\alpha I}$ are complex couplings. Phenomenologically, the HNL is a heavy neutrino, whose mass mixing with active neutrino gives rise to the weak interaction with the coupling suppressed by the mixing angle $U_{\alpha}$. 

HNLs themselves may be responsible for the resolution of all the mentioned problems beyond the SM. In particular, within the scope of neutrino Minimal Standard Model ($\nu$MSM), which adds three HNLs, one with mass in keV range and large lifetime $\tau_{N}\gg t_{\text{Universe}}$, and two quasi-degenerate heavier HNLs, it is possible to resolve all the three phenomena simultaneously (see~\cite{Asaka:2005an,Asaka:2005pn}). The requirement to solve the problems beyond the SM does not fix the HNL mass scale; it may vary from $\mathcal{O}(\text{MeV})$ to $10^{15}\text{ GeV}$ (see~\cite{Chrzaszcz:2019inj,Bondarenko:2021cpc,Klaric:2021cpi,Abdullahi:2022jlv} and references therein). 

HNLs with masses $m_{N}< m_{Z}$ may be copiously produced in decays of heavy bosons $Z$ and $W$, and searched for at the LHC by their decays into charged particles. Such searches are currently running~\cite{CMS:2022fut,ATLAS:2019kpx}, with being significantly improved during the High-Luminosity phase of the LHC (see, e.g.,~\cite{Boiarska:2019jcw}). Future leptonic colliders, such as FCC-ee~\cite{fccphys,fccee} and CEPC~\cite{thecepcstudygroup2018cepc}, that would operate in the regime of $Z$ resonance ($\sqrt{s} = m_{Z}$), may further explore the probed parameter space~\cite{Blondel:2022qqo}, (see~\cite{Blondel:2014bra,Chrzaszcz:2020emg,Shen:2022ffi,Antusch:2016ejd} for the discussion of other type of searches). 

\begin{figure}[!t]
    \centering
    \includegraphics[width=0.9\textwidth]{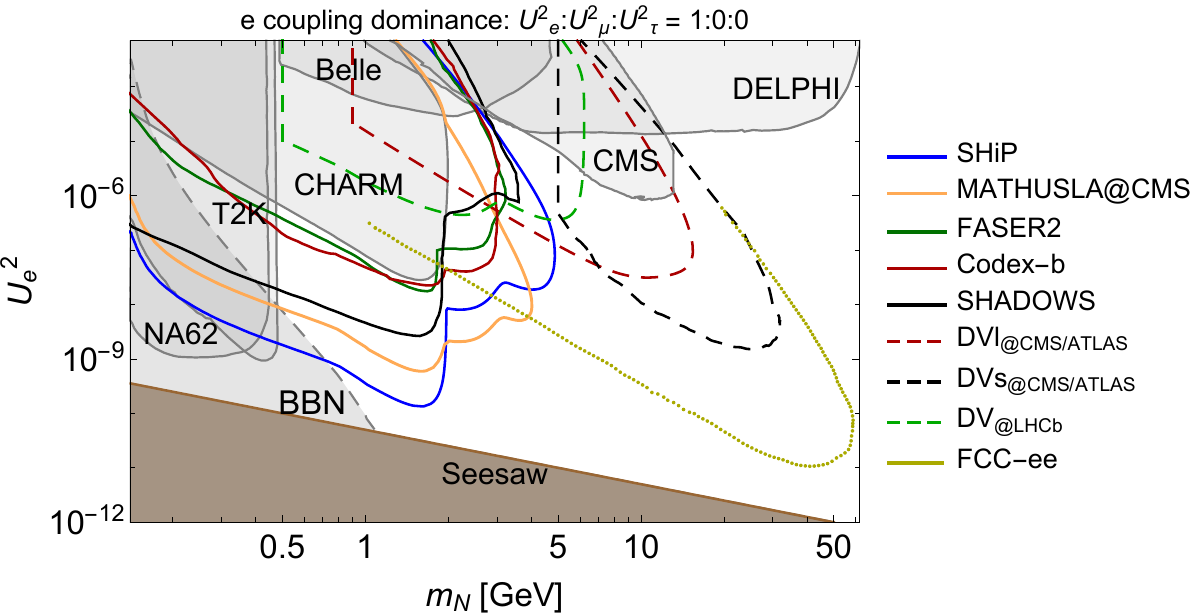}
    \caption{Constraints on heavy neutral leptons which interact solely with $\nu_{e}$. The domain of currently unexplored parameters lies between the regions excluded by past experiments (see~\cite{Abdullahi:2022jlv, Blondel:2022qqo} for references), big bang nucleosynthesis~\cite{Boyarsky:2020dzc, Sabti:2020yrt}, and the region in which couplings of HNLs are too small to generate SM neutrino masses (defined by the seesaw bound). The brown domain indicates the couplings for which HNLs are not able to explain the neutrino oscillations data. The colored curves denote the sensitivity of future proposed experiments, including LHC, LHC-based and extracted beam lines experiments (see references in text), and FCC-ee~\cite{Blondel:2022qqo}.}
    \label{fig:HNL-parameter-space}
\end{figure}

For large masses $m_{N}\simeq 30-60\text{ GeV}$, the reach of FCC-ee is close to the seesaw bound, below which in models with HNLs it is not possible to generate neutrino masses. However, the sensitivity quickly drops with the decrease of the HNL mass, which is because of the scaling of the number of events $N_{\text{events}}\propto N_{Z}\times P_{\text{decay}}$, where the probability to decay inside the detector volume scales as  $P_{\text{decay}} \propto \Gamma_{N}\propto m_{N}^{5}U_{\alpha}^{2}$, see Fig.~\ref{fig:HNL-parameter-space}. 

One of the ways to improve the sensitivity would be to increase the flux of mother particles producing HNLs. During the last decade, a class of such experiments has been proposed to probe the parameter space $m_{N}<m_{B}\simeq 5\text{ GeV}$. These experiments would use $\simeq 3$ orders of magnitude larger flux of mesons $B$, $D$ to produce HNLs. The (incomplete) list includes SHiP~\cite{SHiP:2015vad}, SHADOWS~\cite{Baldini:2021hfw} FASER~\cite{FASER:2018bac,FASER:2018eoc}, MATHUSLA~\cite{MATHUSLA:2018bqv,MATHUSLA:2020uve}, FACET~\cite{Cerci:2021nlb}, CODEX-b~\cite{Aielli:2019ivi}, ANUBIS~\cite{Bauer:2019vqk,Hirsch:2020klk}, AL3X~\cite{2019:AL3X}. One of the strongest potential to probe HNLs with the mixing angles close to the seesaw bound has SHiP. With its reach in the signal~\cite{SHIP:2021tpn} and absence of background (see~\cite{SHIP:2021tpn} and references therein), SHiP is a perfect experiment for the next 10-15 years.

The next step to further explore the parameter space of HNLs would be to search for them at the proposed future hadronic collider FCC-hh~\cite{fcchh}. FCC-hh would collide protons at the energy $\sqrt{s} = 100\text{ TeV}$ with $\simeq 6$ times larger intensity than at the LHC. Thanks to these factors, in particular, the amount of produced $B$ mesons at FCC-hh will be around $30$ times larger that at the LHC. So far, however, the sensitivity of FCC-hh to HNLs from $B$ mesons has not been studied in the literature, see~\cite{Antusch:2016ejd,Curtin:2018mvb,Abdullahi:2022jlv}.

In this paper, we study the sensitivity of FCC-hh to HNLs by considering additional detectors, similar to the proposals of the LHC-based experiments. In order to keep costs of these experiments minimal, we focus on the locations within the infrastructure planned for the FCC. 

The paper is organized as follows. In Sec.~\ref{sec:fcc-hh}, we overview the FCC-hh facility. In Sec.~\ref{sec:fcc-hh-locations}, we discuss various locations of the experiments at FCC-hh. In Sec.~\ref{sec:analytic-estimates}, we study their potential to search for HNLs and compare it with the reach of SHiP by performing analytic estimates and accurate numerical calculations, and make costs estimate for the experiments having the best potential.

\section{FCC-hh}
\label{sec:fcc-hh}

\begin{figure}[!h]
    \centering
    \includegraphics[width=0.7\textwidth]{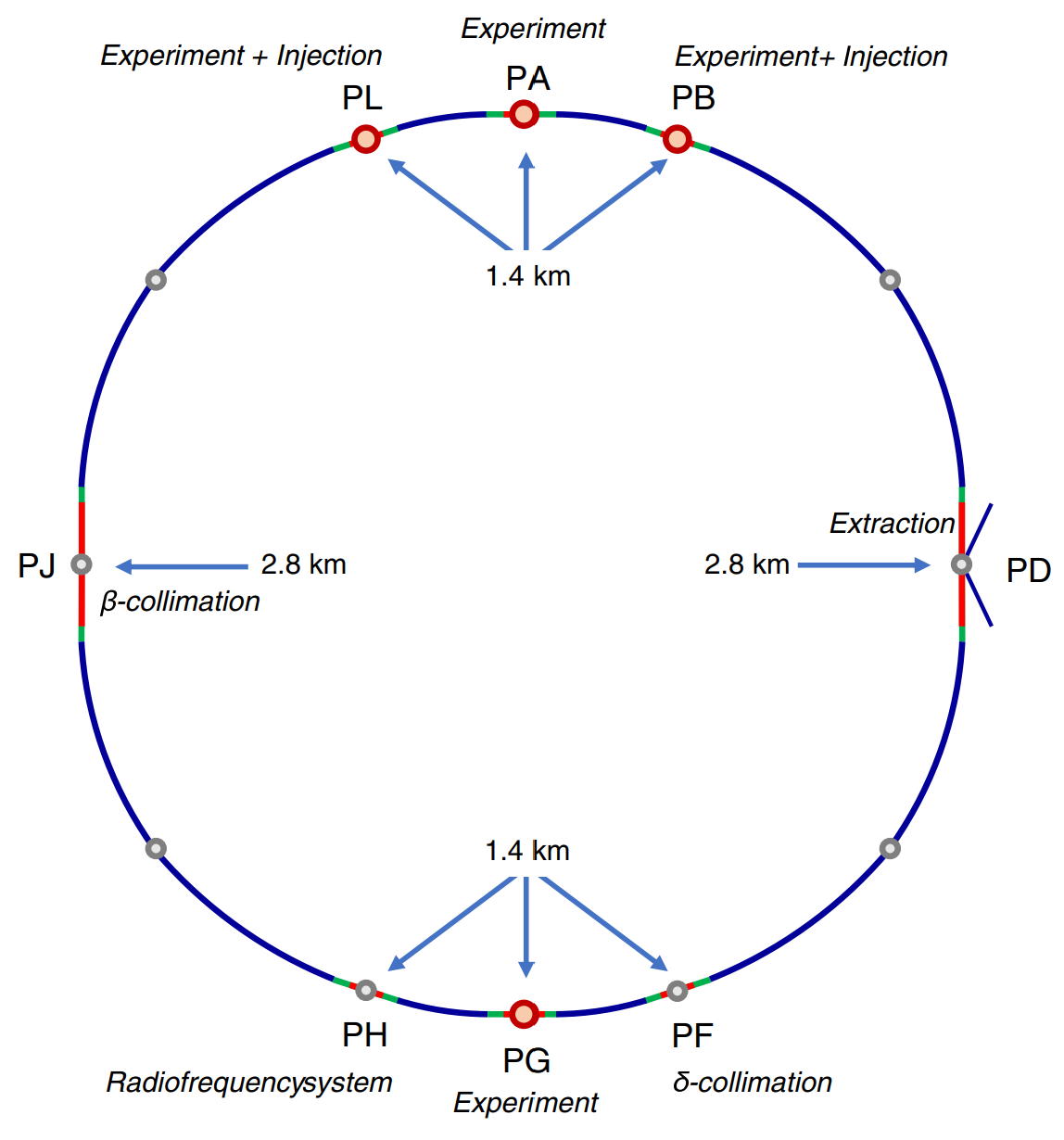}
    \caption{The conceptual layout of FCC-hh. PB, PL denote the proton beams injection points combined with secondary experiments. PA, PG denote the main experiments. PD denotes the beam extraction point. The figure is taken from~\cite{fcchh}.}
    \label{fig:fcc-hh}
\end{figure}
The Future Circular Collider (FCC) will be located at Geneva lake basin and linked to the existing CERN facilities. It has two stages: the electron-positron collider FCC-ee, and the proton-proton collider FCC-hh, with the circumference of around 100 km. The work on the tunnel construction is planned to start around 2030, the first FCC-ee collisions would be around 2040, while FCC-hh would be launched around 2065-2070.

\begin{figure}[!h]
    \centering
    \includegraphics[width = 0.9\textwidth]{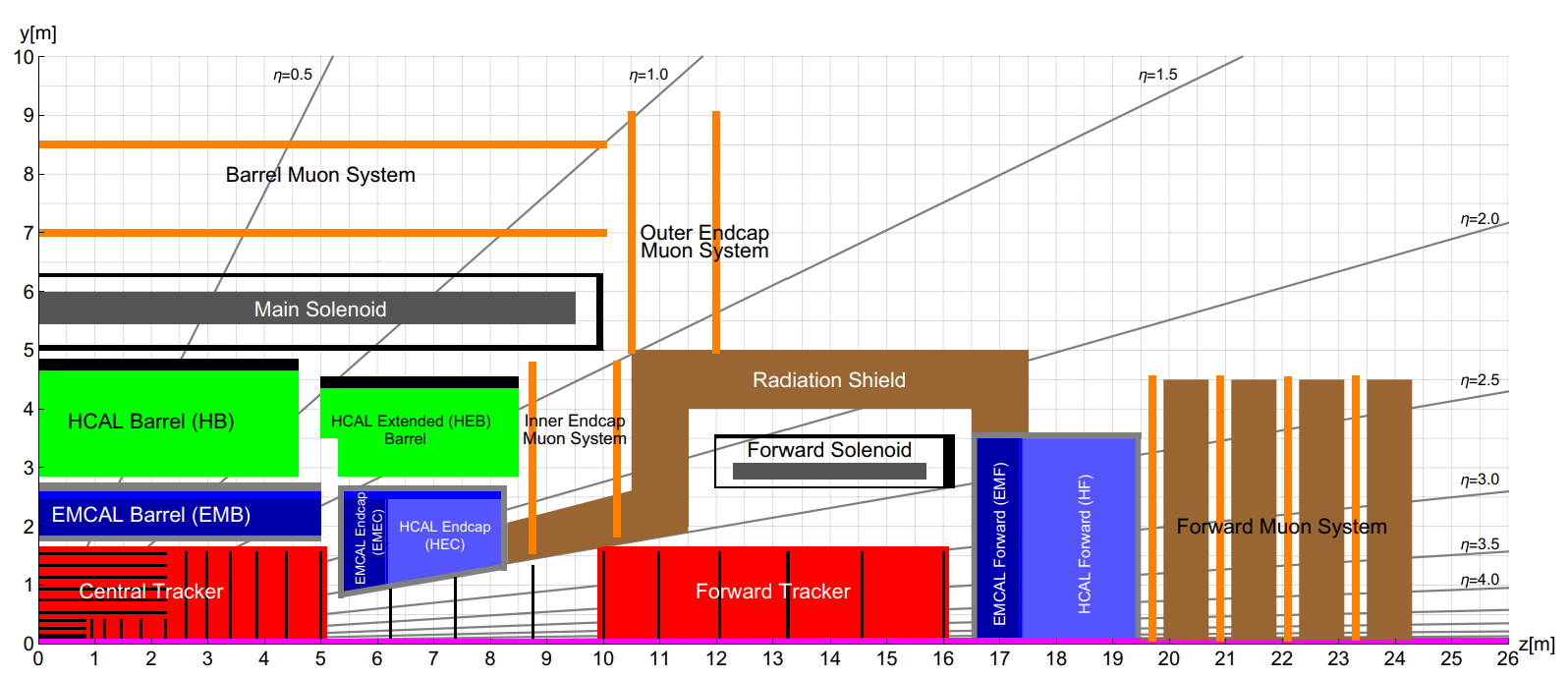}
    \caption{The longitudinal cross section of the FCC-hh reference detector at one of two main experiments. The figure is taken from~\cite{fcchh}.}
    \label{fig:FCCrefdes}
\end{figure}

FCC-hh will continue the scientific program of the LHC. It would operate at the collision energy $\sqrt{s} = 100\text{ TeV}$, accumulating the total luminosity $\mathcal{L} = 20\text{ ab}^{-1}$. The layout of FCC-hh is shown in Fig.~\ref{fig:fcc-hh}. Similarly to the case of the LHC, there are two high-luminosity interaction points PA and PG, and two low-luminosity points PB and PL. Due to larger boosts of particles at FCC-hh compared to the LHC, the main detectors must cover the domain of larger $\eta$ (as compared to $|\eta|<2.5$ at the LHC). The reference detector (see Fig.~\ref{fig:FCCrefdes}) has a diameter of 20 m and length of 50 m. It has the central and forward parts. The central part covers $|\eta|<2.5$ and houses tracking, EM and hadronic calorimetry inside a 4T solenoid. The forward part is displaced from the collision point along the beam axis. It has two forward magnets, resulting in 32 m length of solenoid volume for high-precision momentum spectroscopy up to $|\eta|<4$ and tracking up to $|\eta|<6$.

\section{Experiments at FCC-hh}
\label{sec:fcc-hh-locations}
In this section, we consider possible FCC-based counterparts of the LHC-based experiments. Since we are interested in as low cost spending as possible, the priority will be given to the experiments utilizing the infrastructure planned to construct FCC-hh. We will consider the analogs of FACET and ANUBIS experiments, with modifications caused by different geometry of the infrastructure of FCC-hh. In addition, we discuss an experiment which may be installed in the free space inside the experimental cavern around the main FCC-hh detector. Finally, for illustrative purposes, we also consider a MATHUSLA-like experiment. 

Throughout the paper, we do not study technical aspects of the detectors and, therefore, are only interested in the geometry of the detectors and the corresponding decay volumes. The parameters of the detectors are summarized in Table~\ref{tab:parameters-experiments}.
\begin{table}[!h]
    \centering
    \begin{tabular}{|c|c|c|c|}
    \hline Exp.@FCC & Decay volume dimensions & $\theta_{\text{fid}}$ \\ \hline
    FACET & Annulus, $101<z<119$ m, $r_{\text{in}} = 20\text{ cm},\ r_{\text{out}} = 1\text{ m}$ & (1.6, 10) mrad \\ \hline
     ANUBIS  & Cylinder, $5<z<23\text{ m}$, $R = 9\text{ m}, h = 56\text{ m}$ & (0.8, $\pi/2$) rad \\ \hline
      FCC-hh Mid-$\eta_{1}$ & Annulus, $0<|z|<33$ m, $r_{\text{in}} = 10\text{ m},\ r_{\text{out}} = 15\text{ m}$ & (0.42, $\pi/2$) rad\\ \hline
      FCC-hh Mid-$\eta_{2}$ & Annulus, $12<|z|<33$ m, $r_{\text{in}} = 5\text{ m},\ r_{\text{out}} = 15\text{ m}$ & (0.15, 0.42) rad\\ \hline
      MATHUSLA & Box, $120<y<170\text{ m}$, $x\times z = 200\times 200\text{ m}^{2}$ & (0.44, 0.88) rad \\ \hline 
    \end{tabular}
    \caption{Parameters of FCC-hh-based experiments considered in this paper.}
    \label{tab:parameters-experiments}
\end{table}

\subsection{FACET@FCC}
\begin{figure}[h!]
    \centering
    \includegraphics[height = 5cm]{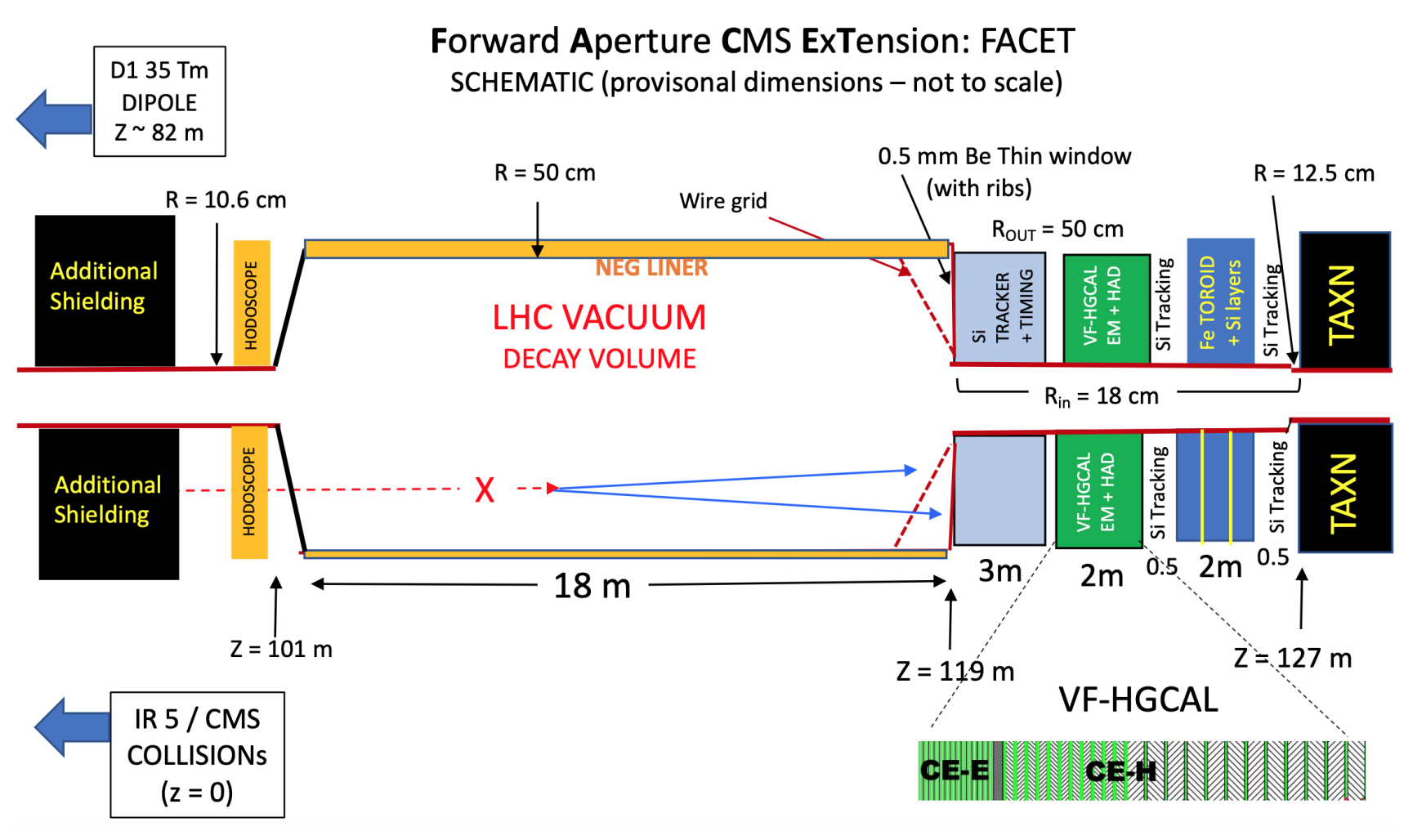}~\includegraphics[height = 5cm]{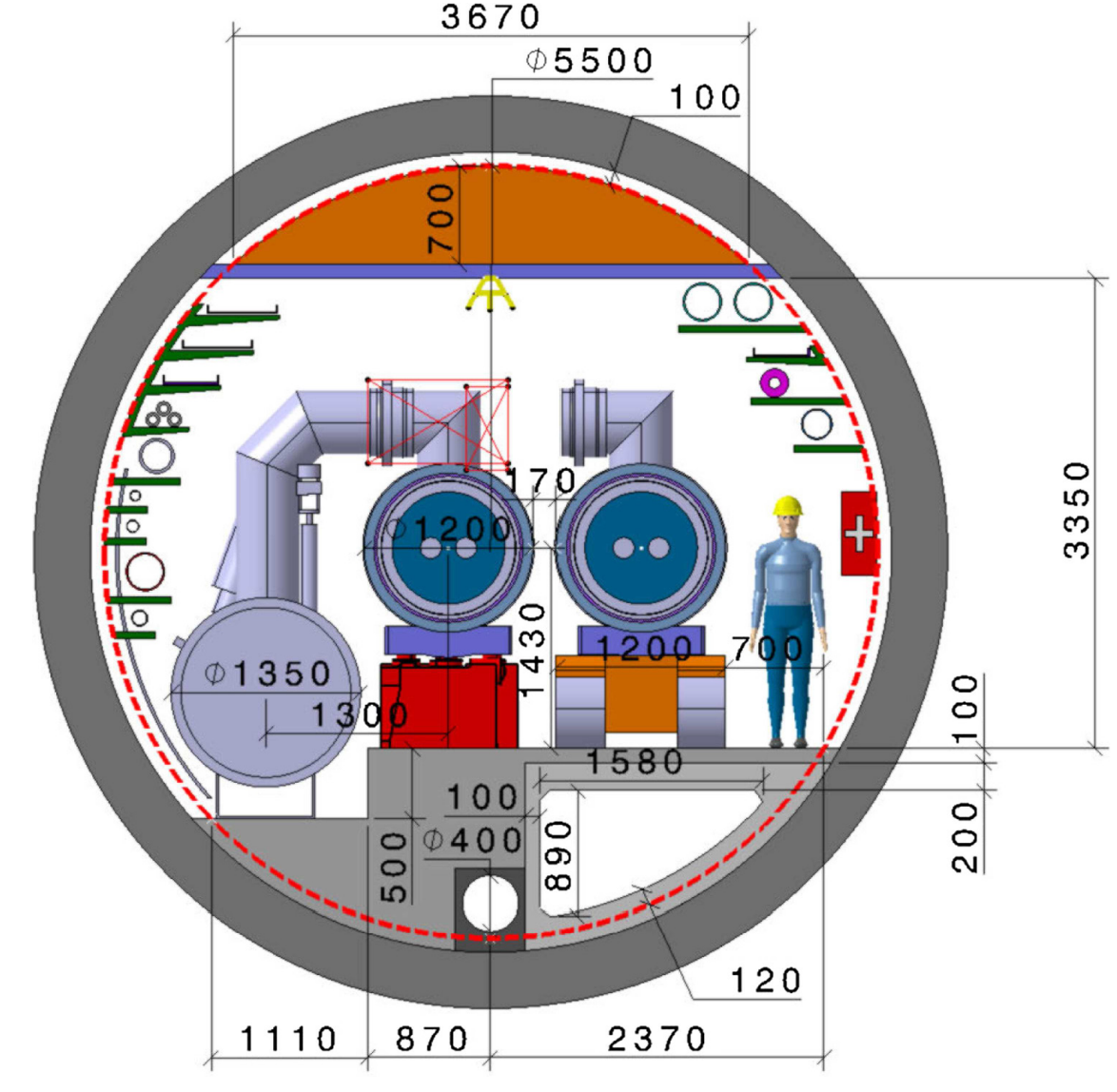}
    \caption{\textit{Left:} Layout of the FACET detector at LHC from~\cite{Cerci:2021nlb}. A section of the proton beam pipe is replaced with a larger pipe followed by detectors. \textit{Right:} cross section of the FCC tunnel~\cite{FCC:2018vvp}. All distances are in mm. The minimal radius of the FACET decay volume is about 20 cm, to account for the beam pipes.}
    \label{fig:FACETlayout}
\end{figure}

FACET is a proposed subsystem of the CMS experiment dedicated to search for FIPs. The concept of the experiment is to replace a section of the beam pipe downstream the interaction point with a larger pipe playing the role of the decay volume, see Fig.~\ref{fig:FACETlayout}. The section is followed by a 8 m long set of detectors surrounding the beam pipe and covering the angular range $\theta \in (1,4)$ mrad. The decay volume is $l_\text{fid} = 18 $ m long, locating from $z=101$ to $z=119$ m. FACET benefits from longer decay volume and closer location as compared to the far-forward experiments (e.g. FASER/FASER2~\cite{FASER:2018bac, FASER:2018eoc}), providing larger solid angle coverage.

We consider a possible modification of the FACET experiment for the FCC-hh --- FACET@FCC. The flux of $B$ mesons produced at FCC-hh is concentrated in the forward direction and, therefore, a forward detector may be suitable for search for FIPs produces in decays of $B$ mesons. On the other hand, forward particles carry large momentum, which suppresses their decay probability $P_{\text{decay}} \approx l_{\text{fid}}/l_{\text{decay}}\propto 1/\gamma$ by the $\gamma$ factor, with $l_\text{decay}$ being the decay length of HNLs. 

We assume that the inner radius of the decay volume is 20 cm in order to envelop the beam pipes. The outer radius is taken to be 1 m, and we assume the detector location to be the same as for FACET, i.e. $z = (101,119)$ m. 

\subsection{ANUBIS@FCC}
The ANUBIS experiment is a proposal to use the shafts near the ATLAS experiment to host a detector to search for FIPs~\cite{Bauer:2019vqk}. Exploiting the existing infrastructure allows to have good detector performance at a moderate cost of the whole experiment. 

\begin{figure}[h!]
    \centering
    \includegraphics[width = 0.3\textwidth]{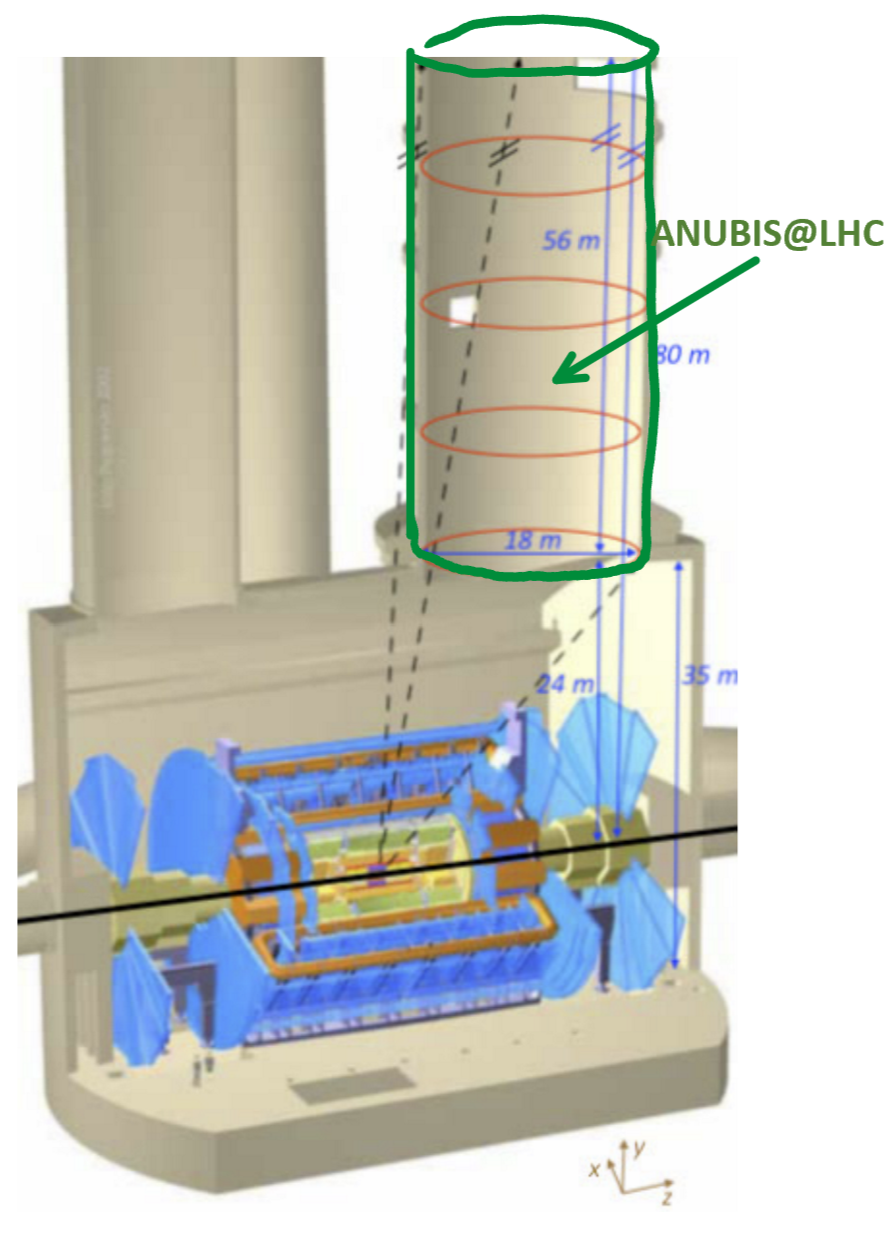}~\includegraphics[width = 0.3\textwidth]{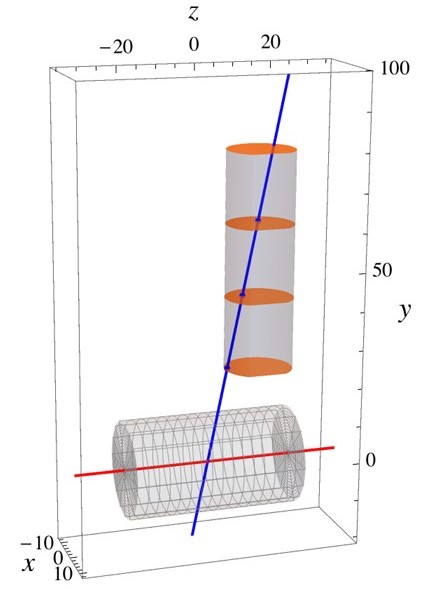}
    
    \includegraphics[width = 0.7\textwidth]{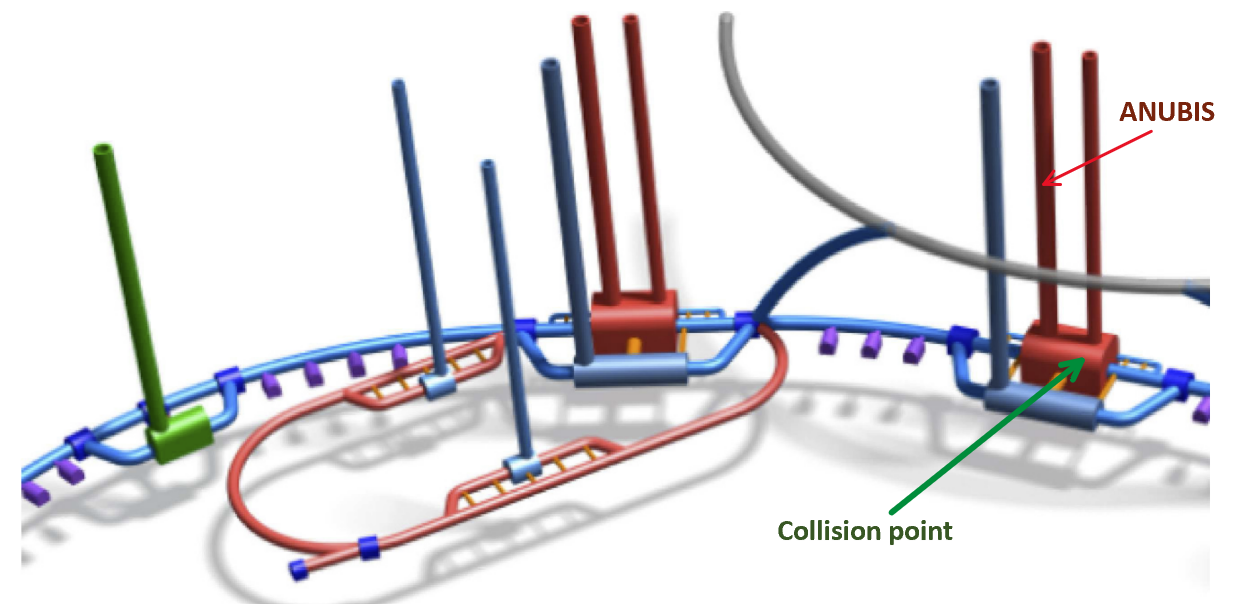}
    \caption{\textit{Top:} detector location and geometry of ANUBIS at LHC in the PX14 shaft near the ATLAS experiment~\cite{Bauer:2019vqk}. Four circular detectors of 9 m radius are separated by about 18.5 m distance in the vertical direction. \textit{Bottom:} possible detector location in the shaft of FCC. }
    \label{fig:ANUBISLHC}
\end{figure}

The detector has the form of a cylinder shown in Fig.~\ref{fig:ANUBISLHC}. Four detector plates of approximately circular shape with 9 m radius are to be installed in the PX14 shaft at equal distances. In the vertical direction, the detector is located at $24\,\text{m}<y<80\,\text{m}$, with the axis of the cylinder being displaced by 14 m in the direction along the beam line. Although the distribution of FIPs at the large angles $\theta\sim 1$ rad covered by the detector is suppressed compared to the forward direction. Nevertheless, it covers much larger solid angle, which may partially compensate for this suppression. In addition, at the lower bound, the decay probability at ANUBIS may be much higher comparable to forward experiments because of the smaller gamma factor and hence increased decay probability of the FIPs. 

To estimate the reach of a similar detector installed at FCC-hh, we assume the same geometry for ANUBIS@FCC.

\subsection{MATHUSLA@FCC}
\begin{figure}[h!]
    \centering
    \includegraphics[height = 4cm]{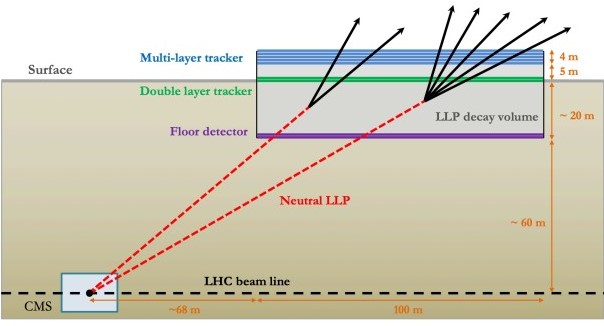}~\includegraphics[height = 4cm]{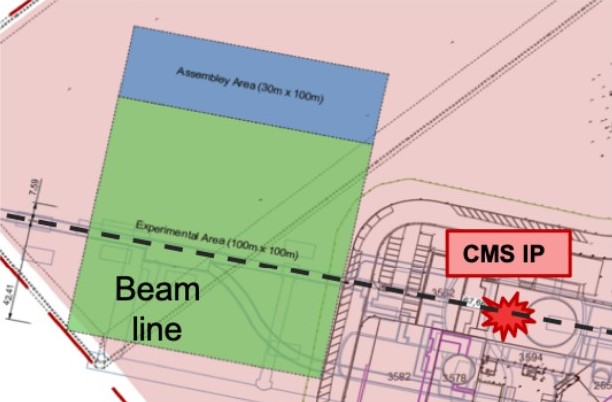}
    \caption{MATHUSLA geometry: side (left) and top (right) views.}
    \label{fig:MATHUSLAlayout}
\end{figure}

MATHUSLA~\cite{MATHUSLA:2020uve,MATHUSLA:2022sze} is a $100\times 100$ m$^2$ box of 25 m height, which is to be located at the surface near the CMS detector. The geometry of the experiment and the top view are shown in Fig.~\ref{fig:MATHUSLAlayout}. The PA interaction point of FCC-hh is planned to be 150 m underground and have a surface site with the size from 6 to 9 hectares~\cite{FCC:2018vvp}. To account for larger displacement and possible improvement of the detector, we consider a simplified setup with the same geometry as shown in Fig.~\ref{fig:MATHUSLAlayout} but with all the distances doubled.

\subsection{Mid-$\eta$}
\label{sec:mid-eta}

The experimental cavern at the FCC-hh interaction point has dimensions $66\times 35\times 35$~m$^3$. The design of the reference detector shown in Fig.~\ref{fig:FCCrefdes} leaves free space inside the cavern, which may be used to host a detector with large angular coverage, such as HECATE~\cite{Chrzaszcz:2020emg}. The decay signature is a displaced vertex inside the decay volume, while the main detector acts as a veto for the SM particles. We propose two options for a new experiment:
\begin{enumerate}
    \item Mid-$\eta_1$: an annulus with the inner and outer radii $10$ and $15$ m, occupying $-33 \text{ m}<z<33 \text{ m}$ and having detector systems at the outer side. This design covers a large solid angle and covers domain of small $\eta$, $|\eta|<1.5$.
    \item Mid-$\eta_2$: two mirrored annuli at $12 \text{ m}<|z|<33\text{ m}$ with the inner 5 m and outer 15 m radii. The detector system is located at the base, in the $x-y$ plane at $|z|=33\text{ m}$. In this case, the detector covers larger $1.5 <|\eta|<2.6$, which allows to have many FIPs even with the smaller solid angle. At the same time, this also implies a larger flux of the SM particles and hence potentially larger background. 
\end{enumerate}

The signal would be the presence of at least two charged tracks that point to a vertex appearing inside the decay volume. We assume the minimal detector configuration allowing to reconstruct vertex with high accuracy without measuring kinematics such as invariant mass and momentum. As for the detector technology, we consider the Resistive Plate Chamber (RPC), and in particular the BIS78 RPCs installed at ATLAS in 2020~\cite{Massa_2020}, which has also been considered in the ANUBIS proposal~\cite{Bauer:2019vqk}. The tracking stations consist of two layers of tracking detectors separated by 1\,m, and have spatial resolution $\delta x \lesssim 1$\,mm, which translates into the angular resolution of order $\delta\alpha \lesssim 0.01$\,rad, while timing resolution $\delta t \lesssim 0.4$\,ns may substantially reduce background. With the total cost of a tracking detector $3.1$k€/m$^2$ and the full detector areas, we estimate the cost of the mid-$\eta_1$ detector as 20\,M€ and mid-$\eta_2$ as 4\,M€. 
\begin{figure}
    \centering
    \includegraphics[width = 0.7\textwidth]{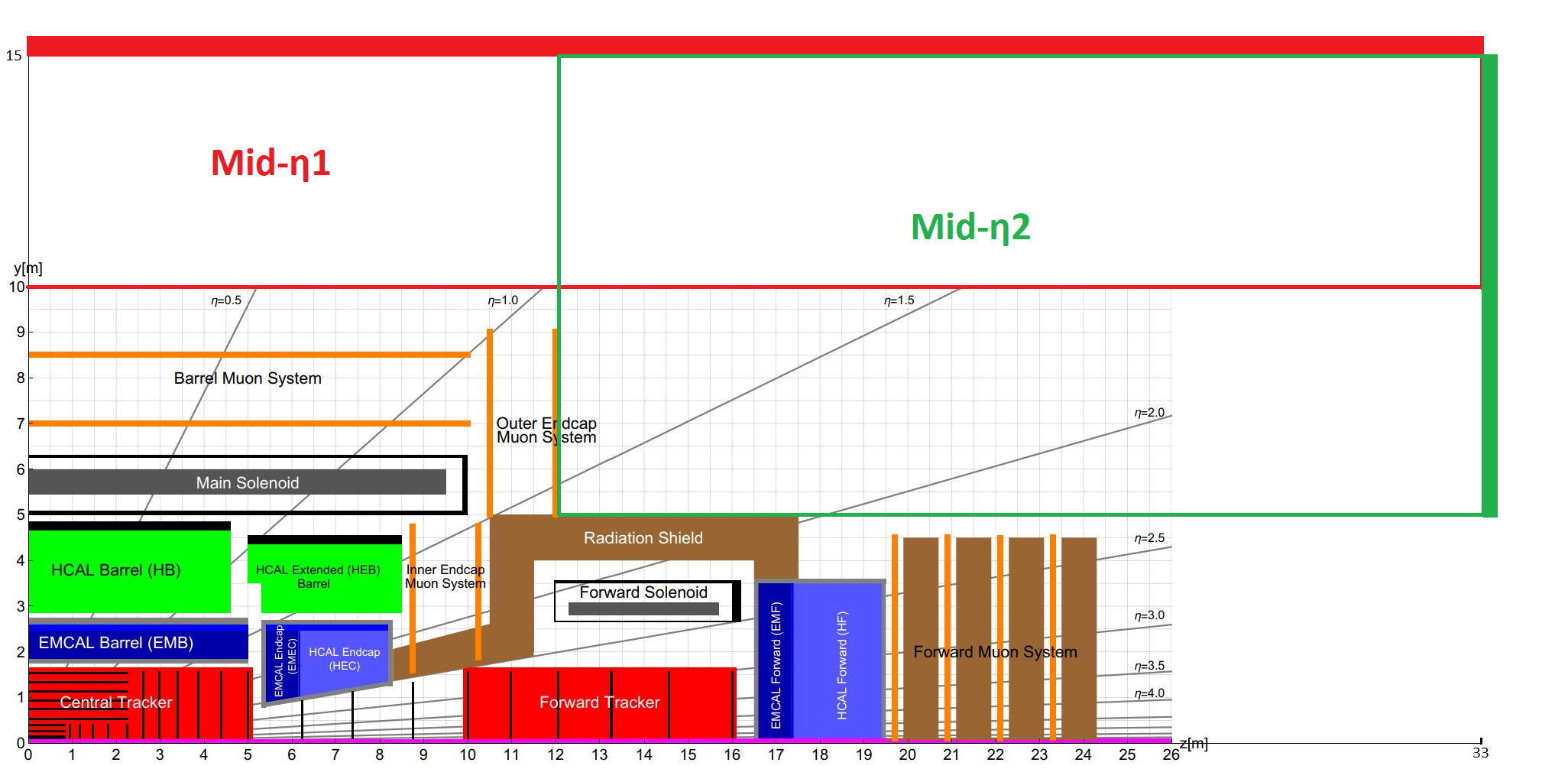}
    \caption{Schematic picture of the geometry of the mid-$\eta$ experiments. The red and green boxes are the decay volumes, while the thick parts represent the detectors. Only the $z>0$ region is shown.}
    \label{fig:twomids}
\end{figure}

\section{Comparison of different locations}
\label{sec:analytic-estimates}
\subsection{Signal yield}
\subsubsection{Analytic estimates}
To make simple estimates of the reaches of the experiments, we follow~\cite{Bondarenko:2019yob}. At the lower bound of the reach, the decay length of a FIP is much larger than the scale of the experiment $l_\text{decay}\gg l_{\text{fid}}$. This allows to find the approximate number of signal events by using
\begin{equation}
\label{eq:simpleest}
    N_{\text{events}} \approx U^4 \times N_B \times \text{Br}_{U^2=1}(B\to N) \times \epsilon_{\text{geom}} \times \frac{m_N}{p_N} \frac{l_\text{fid}}{c\tau_{U^2 = 1}} \times \epsilon_{\text{det}}
\end{equation}
where $N_B = 1.5 \cdot 10^{17}$ is the total number of $B$ mesons at FCC-hh with $20$ ab$^{-1}$ integrated luminosity, $\tau_{U^2 =1} = U^2 \cdot \tau_0$ with $\tau_0$ being the HNL lifetime at rest, $l_\text{fid}$ is the typical length of HNLs paths inside the decay volume, $p_N$ is the average HNL momentum, and  $\text{Br}_{U^2 = 1}(B\to N) = \sum f_{b\to X} \text{Br}(X\to N)/U^2$ is the weighted branching fraction over the meson species with the corresponding values $f_{b\to B^0, B^-}=0.324$, $f_{b\to B_s} = 0.088$, $f_{b\to B_c}= 2.6 \cdot 10^{-3}$. The geometric acceptance $\epsilon_{\text{geom}}$ is the fraction of HNLs whose trajectories lie within the solid angle covered by the detector plates. The detection efficiency $\epsilon_{\text{det}}$ accounts for signal reconstruction efficiency and background suppression. For an estimate, we assume $\epsilon_{\text{det}} = 100\%$.

For simplicity, we approximate the parameters entering Eq.~\eqref{eq:simpleest} by parameters of $B$ mesons, which is a reasonable approximation for heavy HNLs with $m_{N} \simeq m_{B}$. We use the distribution of $B$ mesons at FCC-hh from~\cite{Kling:2021fwx} to compute the averaged values for the experiment-dependent quantities appearing in Eq.~\eqref{eq:simpleest}, which are given in Tab.~\ref{tab:simpleestvalues}.

{\small
\begin{table}[h!]
    \centering
    \begin{tabular}{|c|c|c|c|c|c|c|}
    \hline Exp.@FCC & $\frac{l_{\text{fid}}}{\text{m}}$ & $\epsilon_{\text{geom, }B}$ &$\frac{\langle p_B\rangle}{m_B}$ & $\epsilon_{\text{geom}} \cdot l_\text{fid}\cdot\frac{ m_B}{p_B}$ & $\frac{l_{\text{min}}}{\text{ m}}$& $\frac{\langle p_{B}\rangle}{m_{B}l_{\text{min}}}, \text{m}^{-1}$ \\ \hline
    FACET &18 & 0.08 &150&0.01 &  $1.8$ & $101$ \\ \hline
     ANUBIS  & $\sim 30$ & 0.001 & 1.1& 0.03 & $0.08$ & $\simeq 24$ \\ \hline
      FCC-hh Mid-$\eta_{1}$ & $\gtrsim 5$& 0.16 &1.4&$\gtrsim 0.5$ & $0.2$ & $10$ \\ \hline
      FCC-hh Mid-$\eta_{2}$ & $\sim 20$  &0.17&3.3&0.9 & $0.3$ &$13$\\ \hline
      MATHUSLA & $\sim 70$& 0.007 & 1.6 &0.3 & $0.015$ & 160 \\ \hline
      \multicolumn{6}{|c|}{}
      \\\hline
     MATHUSLA@CMS & 35 & 0.01 & 1.7 & 0.2 & $0.03$ & 80
     \\ \hline
    \end{tabular}
    \caption{Values of the experiment-dependent parameters in Eq.~\eqref{eq:simpleest}: the average length of the decay volume $l_{\text{fid}}$; the geometric acceptance $\epsilon_{\text{geom}}$ -- the fraction of $B$ mesons flying in the direction of detector; the mean momentum of $B$ mesons $p_{B}$; the product of these parameters determining the number of events in the regime of small lifetimes $c\tau \gamma \gg l_{\text{min}}+l_{\text{fid}}$ (larger is better); the characteristic minimal distance from the collision point to the beginning of the decay volume $l_{\text{min}}$; the ratio $\langle p_{B}\rangle/l_{\text{min}}$ determining the sensitiivity in the regime of large lifetimes $c\tau \gamma \lesssim l_{\text{min}}+l_{\text{fid}}$ (larger is better).}
    \label{tab:simpleestvalues}
\end{table}
}

From the table, we see that the mid-$\eta$ experiments have the strongest potential. Given the ratio $N_{B,\text{FCC-hh}}/N_{B,\text{LHC}} \simeq 30$, the number of events at the mid-$\eta$ experiments exceeds the number of events at MATHUSLA@CMS by a factor of 100, which results in an order of magnitude improvement in the sensitivity at the lower bound. 

Compared to these experiments, FACET@FCC has similar $\epsilon_{\text{geom}}$ but much higher $\gamma$ factor. The last feature may improve or worsen the sensitivity, depending on the decay length $l_{\text{decay}} = c\tau \gamma$. In the regime of large lifetimes, when $c\tau$ becomes already comparable to the spatial scale of the experiment, the number of events behaves as $N_{\text{events}}\propto l_{\text{decay}}^{-1} \propto \gamma^{-1}$. This means that FACET@FCC would have worse sensitivity than mid-$\eta$ experiments to very long-lived particles. On the other hand, if the lifetime $c\tau$ is smaller than the distance $l_{\text{min}}$ from the production point to the beginning of the decay volume, one needs large $\gamma$ factor for the decay length to be comparable with $l_{\text{min}}$; otherwise, the particle would decay before reaching the decay volume. The sensitivity of the experiment in this regime improves with the increasing of the ratio $\gamma/l_{\text{min}}$~\cite{Bondarenko:2019yob}, which is the largest for FACET@FCC (see Table~\ref{tab:simpleestvalues}). Therefore, as we will see from the final results, FACET@FCC would have the best sensitivity among the proposed experiments to short-lived particles (Fig.~\ref{fig:sensitivities-comparison}). In addition, unlike the other proposals, FACET is an on-axis experiment, which means that it should have an excellent sensitivity to particles whose production is limited by small polar angles; examples of such models are e.g. dark photon and ALP with the photon coupling~\cite{Beacham:2019nyx}.

ANUBIS has mean $\gamma$ factor comparable to MATHUSLA and mid-$\eta$, but very small geometric acceptance. The sensitivities of ANUBIS@FCC and FACET@FCC are comparable to MATHUSLA@CMS, for which the reduction in $N_{B}$ is compensated by the geometry of the experiment. Finally, MATHUSLA@FCC is competitive with the mid-$\eta$ experiments in the number of events, might have potentially lower background due to the rock shielding, but be very expensive in construction.

\subsubsection{Numeric calculations}
To extend the analytic estimates made in Sec.~\ref{sec:analytic-estimates} by adding the contribution from decays of $D$ mesons and $W$ bosons, estimating the upper bound, and the maximal probed mass, we calculate the sensitivity numerically. The number of events is
\begin{multline}
            N_{\text{events}} \approx \sum_{X = B,W,D}N_{X}\cdot \text{Br}(X\to N)\times \int \dd{\theta_{N}}\dd{l_{N}}\dd{E_{N}} \times \\ \times f^{(X)}_{\theta_{N},E_{N}}\frac{e^{-l_{N}/c\tau_{N}\gamma_{N}}}{c\tau_{N}\gamma_{N}}\cdot\epsilon_{\text{azimuthal}}(l_{N},\theta_{N})\cdot\epsilon_{\text{decay}}(l_{N},\theta_{N},\gamma_{N})\cdot\text{Br}(N\to \text{vis})
        \end{multline}
Here, $f^{(X)}_{\theta_{N},E_{N}}$ is the distribution function of HNLs produced by decays of $X$. $l_{N}$ is the modulus of the path traveled by the HNL from the production point. $\epsilon_{\text{azimuthal}}(l_{N},\theta_{N})$ is the azimuthal acceptance -- the fraction of the azimuthal angle at the given HNL decay point (parametrized by $\theta_{N},l_{N}$) for which the HNL points to the detector plates. $\text{Br}(N\to \text{vis})$ is the branching ratio of decays of HNLs into at least two charged particles.  $\epsilon_{\text{decay}}$ is the decay acceptance -- the fraction of decay products of an HNL with energy $E_{N}$ and decaying at point $\theta_{N},l_{N}$ (within the azimuthal acceptance), which point to the detector plates. 

We use the distributions of $B$ mesons and $D$ mesons at the LHC from FONLL~\cite{Cacciari:1998it,Cacciari:2001td,Cacciari:2012ny,Cacciari:2015fta}, while $W$ bosons at the LHC and particles at FCC-hh from~\cite{Kling:2021fwx}, which used {\sc Pythia}8.\footnote{$B_{c}$ mesons.} For the total numbers of $B,D$ mesons at the LHC, we use the predictions of FONLL at the upper bound of uncertainty (see a discussion on the FONLL predictions in~\cite{Bondarenko:2019yob}). For the total number of $W$ bosons, we use the results of~\cite{ATLAS:2016fij}. For the numbers of particles at FCC-hh, we use the prediction of {\sc Pythia}8. To calculate $f_{\theta_{N},E_{N}}^{X}$, we approximated decays of $X$ by two-body decays into an HNL and electron. Finally, in order to estimate $\epsilon_{\text{decay}}$, we performed a toy simulation, approximating decays of HNLs by three-body decays via the charged current into massless particles.

\begin{figure}[!h]
    \centering
    \includegraphics[width=\textwidth]{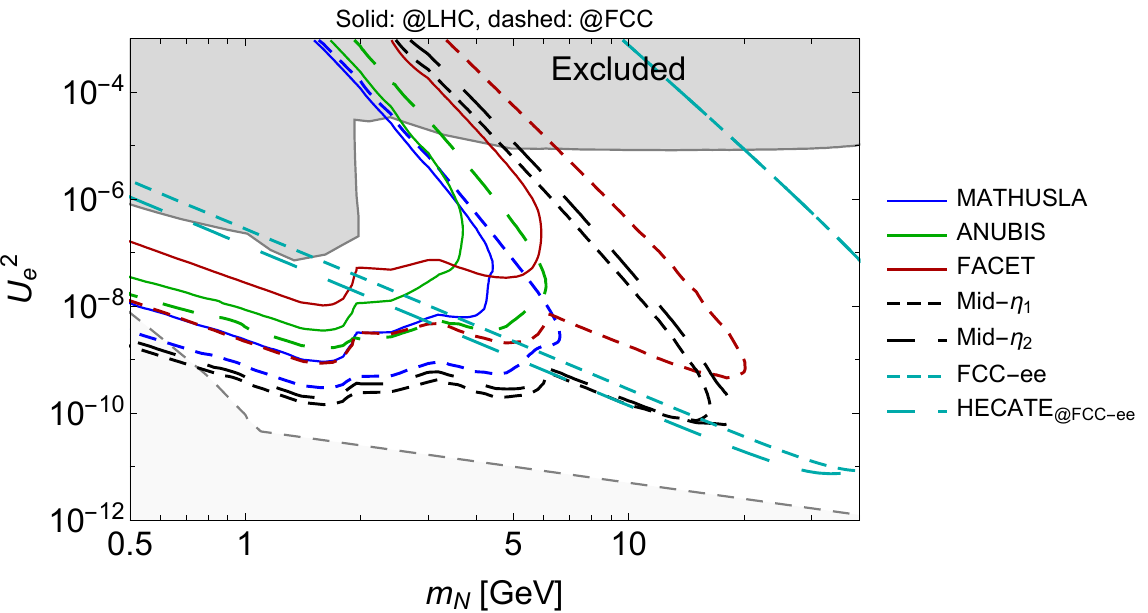}
    \caption{The sensitivity of various LHC-based experiments and experiments at FCC-hh proposed in this paper. For the case of the LHC experiments, we show MATHUSLA, ANUBIS, and (for the first time) FACET, indicated by solid lines; note that the sensitivity of ANUBIS we obtained is significantly weaker than reported in~\cite{Hirsch:2020klk}, see text for details. For the case of the FCC-hh experiments, we show the sensitivities of counterparts of ANUBIS, FACET, and MATHUSLA experiments, as well as the reach of novel proposals -- the mid-$\eta$ experiments, covering $|\eta| < 2.6$ (see Sec.~\ref{sec:fcc-hh-locations}). We also show the sensitivity of FCC-ee and HECATE@FCC-ee, using their setup from~\cite{Abdullahi:2022jlv}. All the curves of the LHC-based and FCC-based experiments are obtained under the assumption of zero background (see Sec.~\ref{sec:backgrounds}) and at 95\% CL.}
    \label{fig:sensitivities-comparison}
\end{figure}

For all experiments, we show the event contours $N_{\text{events}}>3$, which corresponds to 95\% CL region in background-free regime (we will discuss the background in Sec.~\ref{sec:backgrounds}).

\begin{figure}[!h]
    \centering
    \includegraphics[width=0.33\textwidth]{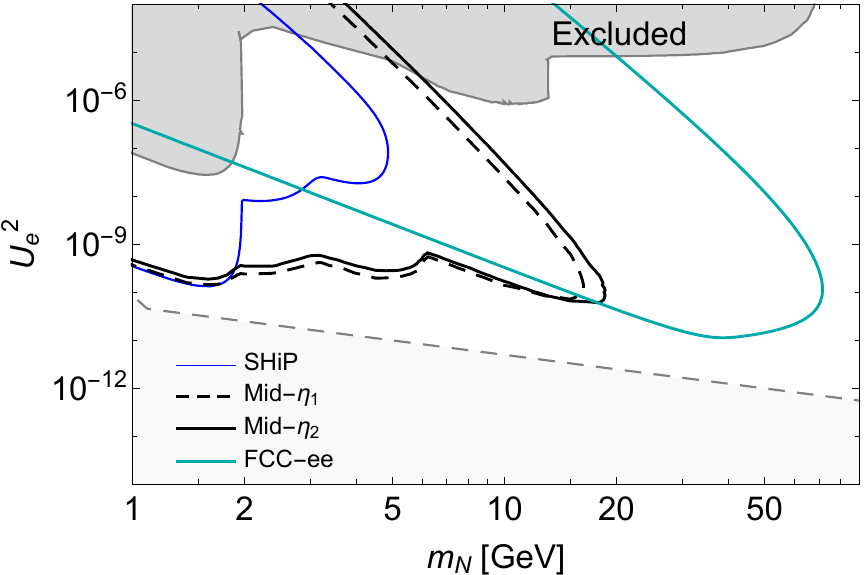}~\includegraphics[width=0.33\textwidth]{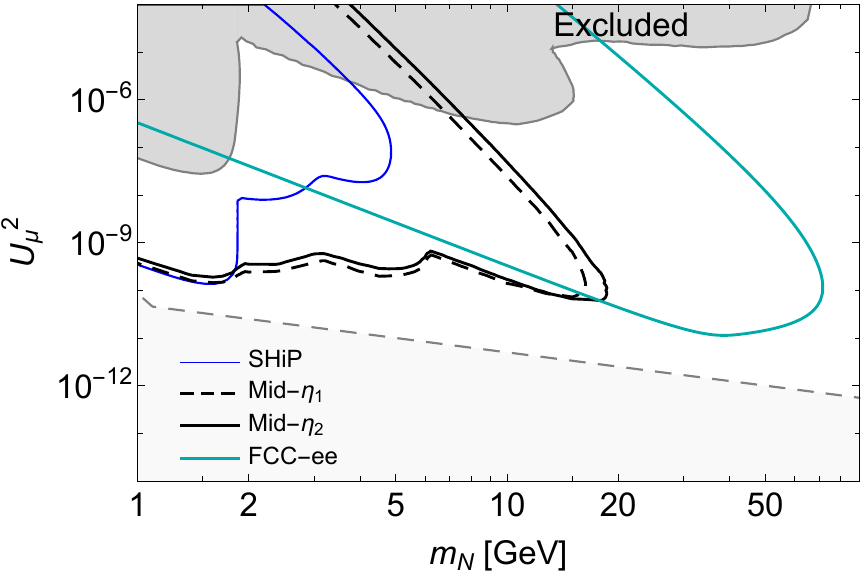}~\includegraphics[width=0.33\textwidth]{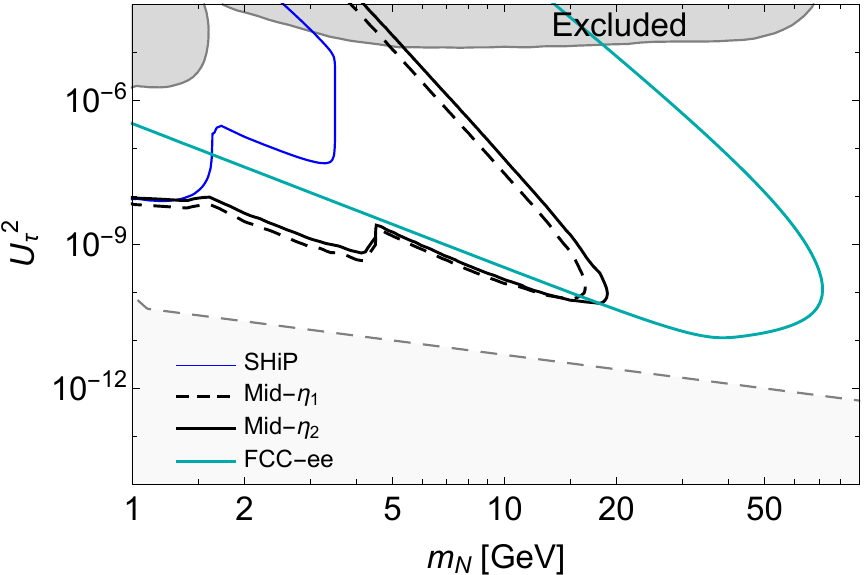}
    \caption{Sensitivity of mid-$\eta$ configurations to HNLs with different mixing patterns: pure $e$, pure $\mu$, pure $\tau$. For reference, the sensitivity regions of FCC-ee and SHiP are also shown. The assumptions are the same as in Fig.~\ref{fig:sensitivities-comparison}.}
    \label{fig:sensitivity-different-flavors}
\end{figure}

The sensitivity of the configurations of experiments at FCC-hh discussed in this paper, together with the sensitivity of MATHUSLA, FACET and ANUBIS at the LHC is shown in Fig.~\ref{fig:sensitivities-comparison}, and for mid-$\eta$ configurations only in Fig.~\ref{fig:sensitivity-different-flavors}. At the lower bound of the sensitivities, the numerical results agree with the analytic estimates. At the upper bound, the scaling agrees with the qualitative estimate $U^{2}_{\text{upper}} \propto \langle E_{N}\rangle/l_{\text{min}}$, where $l_{\text{min}}$ is the minimal distance to the decay volume. 

Together with FACET@FCC, the mid-$\eta$ experiments would provide the opportunity to search for HNLs with mass up to $m_{N}\simeq 20\text{ GeV}$ produced by decays of $W$ bosons. In addition, at the lower bound, the sensitivity of the mid-$\eta$ experiments to HNLs from $W$ bosons may be competitive to the sensitivity of FCC-ee. Indeed, the ratio of the number of events is
\begin{equation}
    \frac{N_{\text{events}}^{W,\text{FCC-hh}}}{N_{\text{events}}^{Z,\text{FCC-ee}}} \sim \frac{N_{W,\text{FCC-hh}}\cdot \epsilon_{\text{geom}}^{(W)}}{N_{Z,\text{FCC-ee}}}\times \frac{\text{Br}(W\to N)}{\text{Br}(Z\to N)}\times \frac{l_{\text{fid}}^{\text{mid-}\eta}}{l_{\text{fig}}^{\text{FCC-ee}}}\times \frac{\langle E_{N}^{\text{FCC-ee}}\rangle}{\langle E_{N}^{\text{mid-}\eta}\rangle}
\end{equation}

Using the mid-$\eta_{2}$ experiment, for masses $m_{N}\lesssim 20\text{ GeV}$ one has $N_{\text{events}}^{W,\text{FCC-hh}}\approx 2.5\cdot 10^{13}$, $\epsilon_{\text{geom}}^{(W)}\approx 0.1$, $\langle E_{N}\rangle \approx 150\text{ GeV}$. Collecting these numbers, one has
\begin{equation}
    \frac{N_{\text{events}}^{W,\text{FCC-hh}}}{N_{\text{events}}^{Z,\text{FCC-ee}}} \sim 0.5 \times \mathcal{O}(1)\times 20 \times 0.2 \times  \frac{\epsilon^{\text{FCC-hh}}_{\text{det}}}{\epsilon^{\text{FCC-ee}}_{\text{det}}}\simeq 2
    \label{eq:FCC-ee-vs-FCC-hh}
\end{equation}
From Eq.~\eqref{eq:FCC-ee-vs-FCC-hh}, we conclude that FCC-hh-based experiments may have better potential to probe small couplings than FCC-ee. However, an important remark should be made with regards to this statement. Namely, FCC-ee would be equipped by the detector system allowing to reconstruct the full kinematics of the decay process, e.g. 4-momenta of the decay products and hence the invariant mass of the decaying particle~\cite{Blondel:2022qqo}. This option is difficult to implement for the most of the FCC-hh-based experiments proposed in this paper (except for FACET@FCC) due to their placement and a large decay volume.\footnote{The same statement is true when comparing FCC-ee and FCC-ee-based experiments, e.g. HECATE.} 

\subsection{Backgrounds}
\label{sec:backgrounds}
Depending on the location of the experiment, the background may consist of: combinatorial muon background; background originated from inelastic scatterings of muons off the decay volume walls; scattering of neutral particles such as neutrons, $K^{0}_{L}$ off air inside the decay volume or their decays; beam-induced background (the beam gas and beam-collimator collisions); cosmic background.

The cosmic background may be removed by the directional cuts; in addition, timing information may be used to disentangle it from the events originated from $pp$ collisions at FCC-hh. For the other backgrounds, FCC-hh may use the methods developed for the LHC-based experiments, keeping the background under control by a combination of various factors: a magnetized iron deflecting charged particles and an high-precision upstream hodoscope for FACET~\cite{Cerci:2021nlb}, hundred meters of rock and a floor veto layer at MATHUSLA~\cite{MATHUSLA:2018bqv,MATHUSLA:2020uve}, off-axis placement and making use of the muon spectrometer and calorimeter at ATLAS as an active veto systems for the muons and neutral hadrons at ANUBIS~\cite{Bauer:2019vqk,Hirsch:2020klk}.

The same methods may be applied to the considered FCC-hh-based experiments, given their similar placements. Let us however separately comment on the mid-$\eta$ experiments, which do not have an analog among the proposed LHC-based experiments. In a close analogy to ANUBIS, the backgrounds may originate from muons and residual hadrons, e.g. $K^{0}_{L}$ and neutrons. These backgrounds, however, would be larger than at ANUBIS due to smaller angles covered by the decay volume. To veto them, one may use the CMS main detector: the muon system~\cite{CMS:2021juv}, and hadronic calorimeter. Finally, a floor veto would provide additional rejecting power.

In general, the conclusion whether the proposed experiments at FCC-hh may be made background-free requires however full-scale simulations, as the background at the FCC-hh-based experiments (even if assuming exactly the same location and geometry as for their LHC analogs) would be more numerous (due to a larger intensity) and more energetic (due to a larger collision energy). 

\section{Conclusions}
Heavy Neutral Leptons are a promising Standard Model extension that may resolve major problems beyond the Standard Model.

It may be possible to probe the parameter space of HNLs in a broad mass range from sub-GeV to a few hundred of GeV at colliders: in the next decade at the LHC in the high luminosity phase, and in a few decades at proposed future colliders, including lepton (FCC-ee, ILC, CEPC), and hadron (FCC-eh, FCC-hh). The sensitivity of the colliders quickly drops with the HNL mass (see Sec.~\ref{sec:introduction}), which is especially important given the natural lower bound in the parameter space of models with HNLs provided by their ability to generate the neutrino masses (the seesaw bound). In particular, the sensitivity of the lepton colliders is very close to the seesaw bound for the HNL mass range $m_{N}\simeq 40-50\text{ GeV}$, but quickly departs from it at smaller masses (Fig.~\ref{fig:HNL-parameter-space}).

\begin{figure}[!h]
    \centering
    \includegraphics[width=0.9\textwidth]{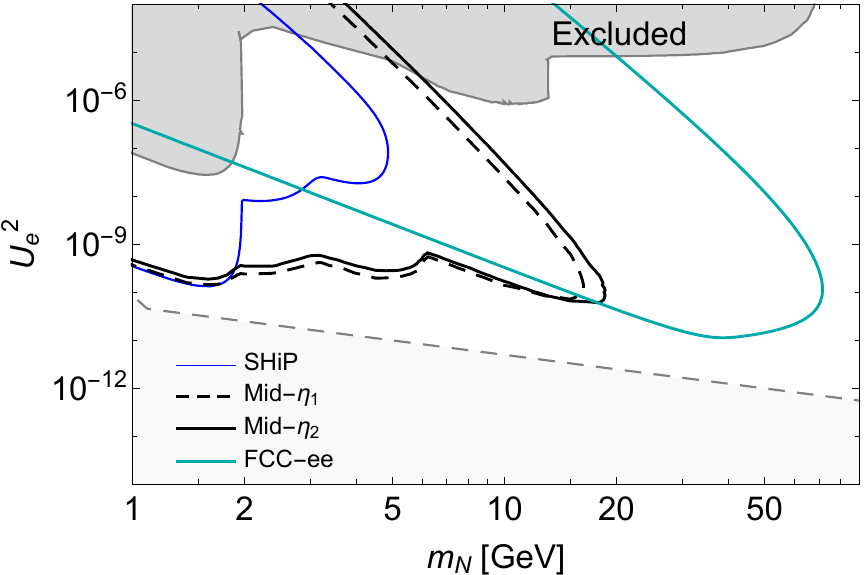}
    \caption{The sensitivity of future proposed experiments to HNLs mixed solely with $\nu_{e}$: SHiP (from~\cite{SHiP:2018yqc}), FCC-ee, and FCC-hh-based experiments proposed in this paper (see Sec.~\ref{sec:mid-eta}): mid-$\eta_{1}$, which covers the domain $|\eta|<1.5$, and mid-$\eta_{2}$, which covers $1.5<|\eta|<2.6$. The curves are obtained under the assumption of zero background has been assumed (see Sec.~\ref{sec:backgrounds}) and at 95\% CL.}
    \label{fig:results}
\end{figure}

There are two options to better explore the parameter space of light HNLs: use larger decay volume, or consider production channels providing much more numerous flux of HNLs, compared to the main production channel considered at colliders, $W/Z \to N+X$. In this paper, we have investigated FCC-hh-based experiments -- a large displaced decay volume to be located near the FCC-hh interaction points. This type of experiments may provide simultaneously with these two options, with the production channel from much more numerous $B$ mesons. 

For the prototypes of these experiments, we have considered analogs of the proposals of the LHC-based experiments, motivated by extensive background and technology studies for the latter (Sec.~\ref{sec:fcc-hh-locations} and Table~\ref{tab:parameters-experiments}): MATHUSLA@FCC, FACET@FCC, ANUBIS@FCC. In addition, we came up with an alternative configuration -- the so-called mid-$\eta$ experiments, that have a decay volume covering the pseudorapidity range $|\eta| <2.5$ (Sec.~\ref{sec:mid-eta}), where we have also estimated the total cost for this experiment at the level of 4M\texteuro.

In Sec.~\ref{sec:analytic-estimates}, we have made an analytic comparison of the reach of these experiments by estimating the quantities defining the lower and upper bounds of the sensitivity to HNLs. We have found (Table~\ref{tab:simpleestvalues}) that the mid-$\eta$ experiments have the largest flux of HNLs for small mixing angles, while FACET@FCC the largest flux for large mixing angles, which is dictated by its on-axis placement. Fig.~\ref{fig:sensitivities-comparison} summarizes numeric sensitivity estimates for these experiments. Compared to the LHC-based experiments, we conclude that may be possible to improve the reach of the LHC at the lower bound by at least one order of magnitude. A more accurate conclusion about the sensitivities may be made after performing a careful background studies: although the background reduction at the proposed experiments may be made with the help of the methods similar to the ones used for the LHC-based experiments, the background at FCC-hh is qualitatively different due to larger collision energy (see Sec.~\ref{sec:backgrounds}).

To summarize, we conclude that the FCC-hh-based experiments may be complementary to lepton colliders, covering the domain of the HNL masses $m_{N}\lesssim 10\text{ GeV}$.

\section*{Acknowledgements}
This project has received funding from the European Research Council (ERC) under the European Union's Horizon 2020 research and innovation programme (GA 694896, GA 758316) and from the NWO Physics Vrij Programme “The Hidden Universe of Weakly Interacting Particles” with project number 680.92.18.03 (NWO Vrije Programma), which is (partly) financed by the Dutch Research Council (NWO).

\newpage

\bibliographystyle{JHEP}
\bibliography{bib.bib}
\end{document}